\documentclass{pasa}%

\usepackage{graphicx}
\input wasyfont.tex
\renewcommand{\deg}{$^\circ$}

\title[The Silurian Hypothesis]{The Silurian Hypothesis: Would it be possible to detect an industrial civilization in the geological record?}
\author[Schmidt and Frank]{Gavin A. Schmidt$^1$ and Adam Frank$^2$
\affil{$^1$NASA Goddard Institute for Space Studies, 2880 Broadway, New York, NY 10025}
\affil{$^2$Department of Physics and Astronomy, University of Rochester, Rochester NY 14620}
}
\jid{in press Int. J. Astrobio}
\doi{10.1017/S1473550418000095}
\jyear{\the\year}

\usepackage{aas_macros}
\usepackage{hyperref} 
\hypersetup{colorlinks,citecolor=blue,linkcolor=blue,urlcolor=blue}

\begin{document}

\begin{frontmatter}
\maketitle
\doublespacing

\begin{abstract}
If an industrial civilization had existed on Earth many millions of years prior to our own era, what 
traces would it have left and would they be detectable today? We summarize the likely geological 
fingerprint of the Anthropocene, and demonstrate that while clear, it will not differ 
greatly in many respects from other known events in the geological record. We then propose tests that 
could plausibly distinguish an industrial cause from an otherwise naturally occurring climate event. 
\end{abstract}

\end{frontmatter}

\section{Introduction}

The search for life elsewhere in the universe is a central occupation of 
astrobiology and scientists have often looked to Earth analogues for 
extremophile bacteria, life under varying climate states and the genesis of 
life itself. A subset of this search is the prospect for intelligent life, 
and then a further subset is the search for civilizations that have the 
potential to communicate with us. A common assumption is that any such 
civilization must have developed industry of some sort.  In particular the 
ability to harness those industrial processes to develop radio technologies 
capable of sending or receiving messages. In what follows, however, we will 
define industrial civilizations here as the ability to harness external 
energy sources at global scales. 

One of the key questions in assessing the likelihood of finding such a 
civilization is an understanding of how often, given that life has arisen 
and that some species are intelligent, does an industrial civilization 
develop? Humans are the only example we know of, and our industrial 
civilization has lasted (so far) roughly 300 years (since, for example, the 
beginning of mass production methods).  This is a small fraction of the time we have existed as a species, and a tiny fraction of the time that complex life has existed on 
the Earth's land surface ($\sim$400 million years ago, Ma). This short 
time period raises the obvious question as to whether this could have 
happened before. We term this the "Silurian Hypothesis"\footnote{We name the 
hypothesis after a 1970 episode of the British science fiction TV series 
Doctor Who where a long buried race of intelligent reptiles "Silurians" are 
awakened by an experimental nuclear reactor. We are not however suggesting 
that intelligent reptiles actually existed in the Silurian age, nor that 
experimental nuclear physics is liable to wake them from hibernation. Other 
authors have dealt with this possibility in various incarnations (for 
instance, \citet{Hogan77}), but it is a rarer theme than we initially assumed.}. 

While much idle speculation and late night chatter has been devoted to this 
question,  we are unaware of previous serious treatments of the  
problem of detectability of prior terrestrial industrial civilizations in
the geologic past.  Given the vast increase in work surrounding exoplanets and questions
related to detection of life, it is worth addressing the question more formally and
in its astrobiological setting.  We note also the recent work of \citet{Wright17} which 
addressed aspects of the problem and previous attempts to 
assess the likelihood of solar system non-terrestrial civilization such as 
\citet{HaqqMisraKumarKopparapu12}. This paper is an attempt to remedy the 
gap in a way that also puts our current impact on the planet into a broader perspective.  
We first note the importance of this question to the well-known Drake 
equation. Then we address the likely geologic consequences of human industrial 
civilization and then compare that fingerprint to potentially similar events 
in the geologic record. Finally, we address some possible research 
directions that might improve the constraints on this question.

\subsection{Relevance to the Drake Equation}

The Drake equation is the well-known framework for estimating of the number of active, 
communicative extraterrestrial civilizations in the Milky Way galaxy \citep{Drake61,Drake65}. 
The number of such civilizations, N, is assumed to be equal to the product of; the 
average rate of star formation, R$^*$, in our galaxy; the fraction of formed stars, $f_p$, that 
have planets; the average number of planets per star, $n_e$, that can potentially support life; 
the fraction of those planets, $f_l$, that actually develop life; the fraction of planets 
bearing life on which intelligent, civilized life, $f_i$, has developed;  the fraction of these 
civilizations that have developed communications, $f_c$, i.e., technologies that release 
detectable signs into space, and the length of time, $L$, over which such civilizations release 
detectable signals.

$$N=R^{\ast }\cdot f_{p}\cdot n_{e}\cdot f_{\ell }\cdot f_{i}\cdot f_{c}\cdot L$$

If over the course of a planet's existence, multiple industrial civilizations can arise over 
the span of time that life exists at all, the value of $f_c$ may in fact be greater than one.  

This is a particularly cogent issue in light of recent developments in astrobiology in which the 
first three terms, which all involve purely astronomical observations, have now been fully determined. 
It is now apparent that most stars harbor families of planets \citep{Seager2013}. Indeed, many of 
those planets will be in the star's habitable zones \citep{Howard2013, DressingCharbonneau2013}. 
These results allow the next three terms to be bracketed in a way that uses the exoplanet data 
to establish a constraint on exo-civilization pessimism.  In  \citet{FrankSullivan2016} such a 
``pessimism line'' was defined as the maximum "biotechnological" probability (per habitable zone 
planets) $f_{bt}$ for humans to be the only time a technological civilization has evolved in cosmic 
history. \citet{FrankSullivan2016} found $f_{bt}$ in the range $\sim$10$^{-24}$ to 10$^{-22}$.

Determination of the "pessimism line" emphasizes the importance of 3 Drake equation terms $f_{\ell }$, 
$f_{i}$ and $f_{c}$. Earth's history often serves as a template for discussions of possible values for 
these probabilities.  For example the has been considerable discussion of how many times life began on 
Earth during the early Archean given the ease of abiogenisis \citep{Patel.et.al15} including the 
possibility of a "shadow biosphere" composed of descendants of a different origin event from the one 
which led to our Last Universal Common Ancestor (LUCA) \citep{ClelandCopley06}.  In addition, there is a 
long standing debate concerning the number of times intelligence has evolved in terms of dolphins and 
other species \citep{Marino15}.  Thus only the term $f_{c}$ has been commonly accepted to have a value on 
Earth of strictly 1. 

\subsection{Relevance to other solar system planets}

Consideration of previous civilizations on other solar system worlds has been taken on by 
\citet{Wright17} and \citet{HaqqMisraKumarKopparapu12}.  We note here that abundant evidence 
exists of surface water in ancient Martian climates (3.8 Ga) 
\citep[e.g.][]{DiAchilleHynek10,Arvidson.et.al14}, and speculation that early Venus (2 Ga to 
0.7 Ga) was habitable (due to a dimmer sun and lower CO$_2$ atmosphere) has been supported by 
recent modeling studies \citep{Way.et.al16}. Conceivably, deep drilling operations could be 
carried out on either planet in future to assess their geological history.  This would constrain
consideration of what the fingerprint might be of life, and even organized civilization 
\citep{HaqqMisraKumarKopparapu12}. Assessments of prior Earth events and consideration of Anthropocene 
markers such as those we carry out below will likely provide a key context for those explorations.  

\subsection{Limitations of the geological record}

That this paper's title question is worth posing is a function of the incompleteness of the 
geological record. For the Quaternary (the last 2.5 million years), there is widespread extant 
physical evidence of, for instance, climate changes, soil horizons, and archaeological evidence
of non-Homo Sapiens cultures (Denisovians, Neanderthals etc.) with occasional evidence of bipedal 
hominids dating back to at least 3.7 Ma (e.g. the Laetoli footprints) \citep{LeakeyHay79}. The oldest
extant large scale surface is in the Negev Desert and is approximately 1.8 Ma old \citep{Matmon.et.al09}. 
However, pre-Quaternary land-evidence is far sparser, existing mainly in exposed sections, drilling, 
and mining operations. In the ocean sediments, due to the recycling of ocean crust, there only exists
sediment evidence for periods that post-date the Jurassic ($\sim$170 Ma) \citep{ODP801}. 

The fraction of life that gets fossilized is always extremely small and varies widely as a function of time, 
habitat and degree of soft tissue versus hard shells or bones \citep{Behrensmeyer.et.al00}. Fossilization 
rates are very low in tropical, forested environments, but are higher in arid environments and fluvial 
systems. As an example, for all the dinosaurs that ever lived, there are only a few thousand near-
complete specimens, or equivalently only a handful of individual animals across thousands of taxa per 
100,000 years. Given the rate of new discovery of taxa of this age, it is clear that species as 
short-lived as Homo Sapiens (so far) might not be represented in the existing fossil record at all.

The likelihood of objects surviving and being discovered is similarly unlikely. \citet{Zalasiewicz09} 
speculates about preservation of objects or their forms, but the current area of urbanization is less 
than 1\% of the Earth's surface \citep{Schneider.et.al09}, and exposed sections and drilling sites for 
pre-Quaternary surfaces are orders of magnitude less as fractions of the original surface. Note that even 
for early human technology, complex objects are very rarely found. For instance, the Antikythera 
Mechanism (ca. 205 BCE) is a unique object until the Renaissance. Despite impressive recent gains in the 
ability to detect the wider impacts of civilization on landscapes and ecosystems \citep{Kidwell15}, we 
conclude that for potential civilizations older than about 4 Ma, the chances of finding direct evidence 
of their existence via objects or fossilized examples of their population is small. We note, however, that one might ask the indirect question related to antecedents in the fossil record indicating species that might {\it lead} downstream to the evolution of later civilization-building species.  Such arguments, for or against, the Silurian hypothesis would rest on evidence concerning highly social behavior or high intelligence based on brain size. The claim would then be that there are other species in the fossil record which could, or could not, have evolved into civilization-builders.  In this paper, however, we focus on physico-chemical tracers for previous industrial civilizations.  In this way there is an opportunity to widen the search to tracers that are more widespread, even though they may be subject to more varied interpretations.

\subsection{Scope of this paper}

We will restrict the scope of this paper to geochemical constraints on the existence of pre-Quaternary 
industrial civilizations, that may have existed since the rise of complex life on land. This rules out 
societies that might have been highly organized and potentially sophisticated but that did not develop 
industry and probably any purely ocean-based lifeforms. The focus is thus on the period between 
the emergence of complex life on land in the Devonian ($\sim$400 Ma) in the Paleozoic era and the 
mid-Pliocene ($\sim$4 Ma). 

\section{The geological footprint of the Anthropocene}

While an official declaration of the Anthropocene as a unique geological era is still pending 
\citep{Crutzen02, Zalasiewicz.et.al17}, it is already clear that our human efforts will impact the 
geologic record being laid down today \citep{Waters.et.al14}. Some of the discussion of the specific 
boundary that will define this new period is not relevant for our purposes because the markers proposed 
(ice core gas concentrations, short-half-lived radioactivity, the Columbian exchange) \citep[e.g.][]
{LewisMaslin15, Hamilton16} are not going to be geologically stable or distinguishable on multi-million 
year timescales. However, there are multiple changes that have already occurred that will persist. We 
discuss a number of these below.

There is an interesting paradox in considering the Anthropogenic footprint on a geological timescale. The 
longer human civilization lasts, the larger the signal one would expect in the record. However, the 
longer a civilization lasts, the more sustainable its practices would need to have become in order to 
survive. The more sustainable a society (e.g. in energy generation, manufacturing, or agriculture) the 
smaller the footprint on the rest of the planet. But the smaller the footprint, the less of a signal 
will be embedded in the geological record. Thus the footprint of civilization might be self-limiting on 
a relatively short time-scale. To avoid speculating about the ultimate fate of humanity, we will 
consider impacts that are already clear, or that are foreseeable under plausible trajectories for the 
next century \citep[e.g.][]{Kohler16,Nazarenko.et.al15}. 

We note that effective sedimentation rates in ocean sediment for cores with multi-million-year old 
sediment are on the order of a few cm/1000 years at best, and while the degree of bioturbation may smear 
a short period  signal, the Anthropocene will likely only appear as a section a few cm thick, and 
appear almost instantaneously in the record.

\subsection{Stable isotope anomalies of carbon, oxygen, hydrogen and nitrogen}

Since the mid-18th Century, humans have released over 0.5 trillion tons of fossil carbon via the 
burning of coal, oil and natural gas \citep{LeQuere.et.al16}, at a rate orders of magnitude 
faster than natural long-term sources or sinks. In addition, there has been widespread 
deforestation and addition of carbon dioxide into the air via biomass burning. All of this carbon 
is biological in origin and is thus depleted in $^{13}$C compared to the much larger pool of 
inorganic carbon \citep{RevelleSuess57}. Thus the ratio of $^{13}$C to $^{12}$C in the 
atmosphere, ocean and soils is decreasing (an impact known as the ``Suess Effect'' 
\citep{Quay.et.al92}) with a current change of around -1\permil\ $\delta^{13}$C 
since the pre-industrial \citep{Bohm.et.al02,Eide.et.al17} in the surface ocean and atmosphere (figure~\ref{stable}a).  

As a function of the increase of fossil carbon into the system, augmented by black carbon 
changes, other non-CO$_2$ trace greenhouse gases (like N$_2$O, CH$_4$ and chloro-fluoro-carbons (CFCs)), global 
industrialization has been accompanied by a warming of about 1\deg C so far since the mid 19th 
Century \citep{GISTEMP,IPCCAR5_Ch10}. Due to the temperature-related fractionation in the formation of 
carbonates \citep{KimONeil97} (-0.2\permil\ $\delta^{18}$O per \deg C) and strong correlation in 
the extra-tropics between temperature and $\delta^{18}$O (between 0.4 and 0.7\permil\ per \deg C) 
(and roughly 8$\times$ as sensitive for deuterium isotopes relative to hydrogen ($\delta$D)), we 
expect this temperature rise to be detectable in surface ocean carbonates (notably foraminifera), 
organic biomarkers, cave records (stalactites), lake ostracods and high-latitude ice cores, 
though only the first two of these will be retrievable in the time-scales considered here.

The combustion of fossil fuel, the invention of the Haber-Bosch process, the large-scale 
application of nitrogenous fertilizers, and the enhanced nitrogen fixation associated with 
cultivated plants, have caused a profound impact on nitrogen cycling \citep{Canfield.et.al10}, 
such that $\delta^{15}$N anomalies are already detectable in sediments remote from civilization 
\citep{Holtgrieve.et.al11}.

\subsection{Sedimentological records}

There are multiple causes of a greatly increased sediment flow in rivers and therefore in 
deposition in coastal environments. The advent of agriculture and associated deforestation have 
lead to large increases in soil erosion \citep{Goudie00, CommAg10}. Furthermore, canalization of 
rivers (such as the Mississippi) have led to much greater oceanic deposition of sediment than 
would otherwise have occurred. This tendency is mitigated somewhat by concurrent increases in 
river dams which reduce sediment flow downstream. Additionally, increasing temperatures and 
atmospheric water vapor content have led to greater intensity of precipitation \citep{Kunkel.et.al13} 
which, on its own, would also lead to greater erosion, at least regionally. Coastal erosion is 
also on the increase as a function of rising sea level, and in polar regions is being enhanced by 
reductions in sea ice and thawing permafrost \citep{Overeem.et.al11}. 

In addition to changes in the flux of sediment from land to ocean, the composition of the 
sediment will also change. Due to the increased dissolution of CO$_2$ in the ocean as a 
function of anthropogenic CO$_2$ emissions, the upper ocean is acidifying (a 26\% increase 
in H$^+$ or 0.1 pH decrease since the 19th Century) \citep{Orr.et.al05}. This will lead to 
an increase in CaCO$_3$ dissolution within the sediment that will last until the ocean can 
neutralize the increase. There will also be important changes in mineralogy 
\citep{Zalasiewicz.et.al13,Hazen.et.al17}. Increases in continental weathering are also likely to change 
ratios of strontium and osmium (e.g. $^{87}$Sr/$^{86}$Sr and $^{187}$Os/$^{188}$Os ratios) 
\citep{Jenkyns10}. 

As discussed above, nitrogen load in rivers is increasing as a function of agricultural 
practices. This in turn is leading to more microbial activity in the coastal ocean which can 
deplete dissolved oxygen in the water column \citep{DiazRosenberg08}, and recent syntheses 
suggests a global decline already of about 2\% \citep{Schmidtko.et.al17,Ito.et.al17}. This in turn is 
leading to an expansion of the oxygen minimum zones, greater ocean anoxia, and the creation 
of so-called ``dead-zones'' \citep{Breitburg.et.al17}. Sediment within these areas will thus have greater organic 
content and less bioturbation \citep{Tyrrell11}. The ultimate extent of these dead zones is 
unknown. 

Furthermore, anthropogenic fluxes of lead, chromium, antimony, rhenium, platinum group metals, rare earths and 
gold, are now much larger than their natural sources \citep{Galuszka.et.al13,SenPeuckerEhrenbrink12}, 
implying that there will be a spike in fluxes in these metals in river outflow and hence higher 
concentrations in coastal sediments.

\subsection{Faunal radiation and extinctions}

The last few centuries have seen significant changes in the abundance and spread of small 
animals, particularly rats, mice, and cats etc., that are associated with human exploration 
and biotic exchanges. Isolated populations almost everywhere have now been superseded in 
many respects by these invasive species. The fossil record will likely indicate a large 
faunal radiation of these indicator species at this point. Concurrently, many other species 
have already, or are likely to become, extinct, and their disappearance from the fossil 
record will be noticeable. Given the perspective from many million years ahead, large mammal 
extinctions that occurred at the end of the last ice age will also associated with the onset 
of the Anthropocene. 

\subsection{Non-naturally occurring synthetics}

There are many chemicals that have been (or were) manufactured industrially that for various 
reasons can spread and persist in the environment for a long time \citep{Bernhardt.et.al17}. Most 
notably, persistent organic pollutants (POPs) (organic molecules that are resistant to degradation by 
chemical, photo-chemical or biological processes), are known to have spread across the world (even to 
otherwise pristine environments) \citep{Beyer.et.al00}. Their persistence is often tied to being 
halogenated organics since the bond strength of C-Cl (for instance) is much stronger than C-C. For 
instance, polychlorinated biphenyls (PCBs) are known to have lifetimes of many hundreds of 
years in river sediment \citep{Bopp79}. How long a detectable signal would persist in ocean sediment is, 
however, unclear. 

Other chlorinated compounds may also have the potential for long-term preservation, 
specifically CFCs and related compounds. While there are natural 
sources for the most stable compound (CF$_4$), there are only anthropogenic sources for 
C$_2$F$_6$ and SF$_6$, the next most stable compounds. In the atmosphere, their sink via 
photolytic destruction in the stratosphere limits their lifetimes to a few thousand years 
\citep{Ravishankara.et.al93}. The compounds do dissolve in the the ocean and can be used as 
tracers of ocean circulation, but we are unaware of studies indicating how long these 
chemicals might survive and/or be detectable in ocean sediment given some limited evidence for microbial 
degradation in anaerobic environments \citep{DenovanStrand92}.

Other classes of synthetic biomarkers may also persist in sediments. For instance, steroids, leaf waxes, 
alkenones, and lipids can be preserved in sediment for many millions of years 
\citep[i.e][]{Pagani.et.al06}. What might distinguish naturally occurring biomarkers from synthetics 
might be the chirality of the molecules. Most total synthesis pathways do not discriminate between 
D- and L- chirality, while biological processes are almost exclusively monochiral 
\citep{Meierhenrich08} (for instance, naturally occurring amino acids are all L-forms, and 
almost all sugars are D-forms). Synthetic steroids that do not have natural counterparts are also now 
ubiquitous in water bodies.

\subsection{Plastics}

Since 1950, there has been a huge increase in plastics being delivered into the ocean 
\citep{Moore08, Eriksen.et.al14}. Although many common forms of plastic (such as polyethylene and
polypropylene) are buoyant in sea water, and even those that are nominally heavier than water may be 
incorporated into flotsam that remains at the surface,  it is already clear that mechanical erosional 
processes will lead to the production of large amounts of plastic micro and nano-particles 
\citep{Cozar.et.al14, Andrady15}. Surveys have shown increasing amounts of plastic `marine litter' on the 
seafloor from coastal areas to deep basins and the Arctic \citep{Pham.et.al14, Tekman.et.al17}. On
beaches, novel aggregates ``plastiglomerates'' have been found where plastic-containing debris comes 
into contact with high temperatures \citep{Corcoran.et.al14}. 

The degradation of plastics is mostly by solar ultra-
violet radiation and in the oceans occurs mostly in the photic zone \citep{Andrady15} and is notably 
temperature dependent \citep{Andrady.et.al98} (other mechanisms such as thermo-oxidation or 
hydrolysis do not readily occur in the ocean). The densification of small plastic particles by fouling 
organisms, ingestion and incorporation into organic `rains` that sink to the sea floor is an effective 
delivery mechanism to the seafloor, leading to increasing accumulation in ocean sediment where 
degradation rates are much slower \citep{Andrady15}. Once in the sediment, microbial activity is a 
possible degradation pathway \citep{Shah.et.al08} but rates are sensitive to oxygen availability and 
suitable microbial communities. 

As above, the ultimate long-term fate of these plastics in sediment is unclear, but the potential for 
very long term persistence and detectability is high. 

\subsection{Transuranic elements}

Many radioactive isotopes that are related to anthropogenic fission or nuclear arms, have 
half-lives that are long, but not long enough to be relevant here. However, there are two 
isotopes that are potentially long-lived enough. Specifically, Plutonium-244 (half-life 80.8 
million years) and Curium-247 (half-life 15 million years) would be detectable for a large 
fraction of the relevant time period if they were deposited in sufficient quantities, say, 
as a result of a nuclear weapon exchange. There are no known natural sources of $^{244}$Pu 
outside of supernovae. 

Attempts have been made to detect primordial $^{244}$Pu on Earth with mixed success 
\citep{Hoffman.et.al71,Lachner.et.al12}, indicating the rate of actinide meteorite accretion 
is small enough \citep{Wallner.et.al15} for this to be a valid marker in the event of a 
sufficiently large nuclear exchange. Similarly, $^{247}$Cm is present in nuclear fuel waste 
and as a consequence of a nuclear explosion. 

Anomalous isotopic ratios in elements with long-lived radioactive isotopes are also possible 
signatures, for instance, lower than usual $^{235}$U ratios, and the presence of expected daughter 
products, in uranium ores in the Franceville Basin in the Gabon have been traced to naturally occurring 
nuclear fission in oxygenated, hydrated rocks around 2 Ga \citep{GauthierLafaye.et.al96}. 

\begin{figure}[h]
\includegraphics[width=\columnwidth/2]{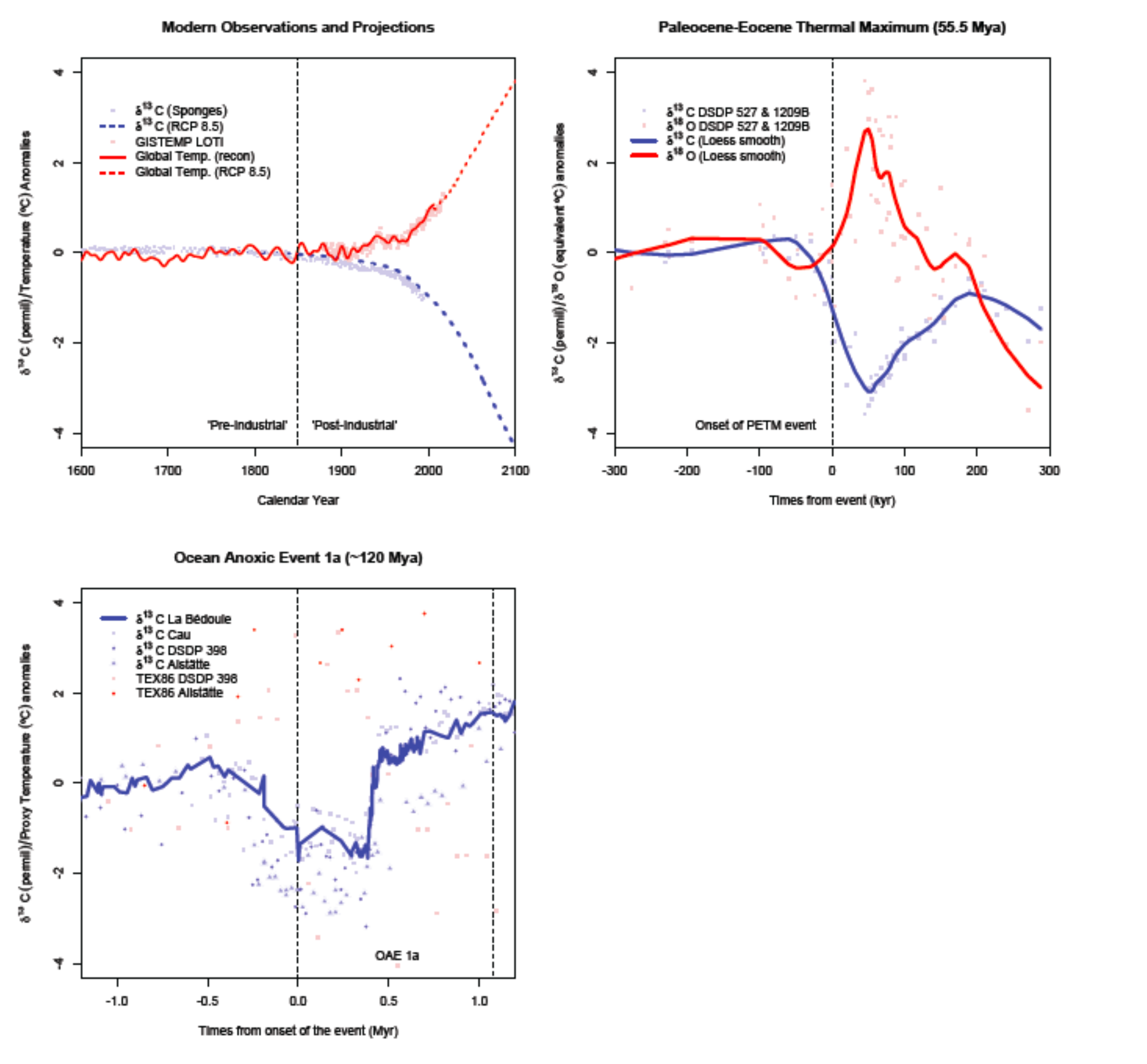}
\caption{Illustrative stable carbon isotopes and temperature (or proxy) profiles across three periods. a) The modern era (from 1600 CE with projections to 2100). Carbon isotopes are from sea sponges \citep{Bohm.et.al02}, and projections from \citet{Kohler16}. Temperatures are from \citet{Mann.et.al08} (reconstructions), GISTEMP \citep{Hansen.et.al10} (instrumental) and projected to 2100 using results from \citet{Nazarenko.et.al15}. Projections assume trajectories of emissions associated with RCP8.5 \citep{vanVuuren.et.al11}. b) The Paleocene-Eocene Thermal Maximum (55.5 Ma). Data from two DSDP cores (589 and 1209B) \citep{TripatiElderfield04} are used to estimate anomalous isotopic changes and a loess smooth with a span of ~200 kya is applied to make the trends clearer. Temperatures changes are estimated from observed $\delta^{18}$O$_{carbonate}$ using a standard calibration \citep{KimONeil97}. c) Oceanic Anoxic Event 1a (about 120 Ma). Carbon isotopes are from the La B\'edoule and Cau cores from the paleo-Tethys \citep{Kuhnt.et.al11, Naafs.et.al16} aligned as in \citet{Naafs.et.al16} and placed on an approximate age model. Data from Alst\"atte \citep{BottiniMutterlose12} and DSDP Site 398 \citep{Li.et.al08} are aligned based on coherence of the $\delta^{13}$C anomalies. Temperature change estimates are derived from TEX86 \citep{Mutterlose.et.al14,Naafs.et.al16}. Note that the y-axis spans the same range in all three cases, while the timescales vary significantly.}
\label{stable}
\end{figure}

\subsection{Summary}

The Anthropocene layer in ocean sediment will be abrupt and multi-variate, consisting of 
seemingly concurrent specific peaks in multiple geochemical proxies, biomarkers, elemental composition, 
and mineralogy. It will likely demarcate a clear transition of faunal taxa prior to the event compared 
to afterwards. Most of the individual markers will not be unique in the context of Earth 
history as we demonstrate below, but the combination of tracers may be. However, we speculate 
that some specific tracers that would be unique, specifically persistent synthetic 
molecules, plastics, and (potentially) very long-lived radioactive fallout in the event of nuclear 
catastrophe. Absent those markers, the uniqueness of the event 
may well be seen in the multitude of relatively independent fingerprints as opposed to a 
coherent set of changes associated with a single geophysical cause.

\section{Abrupt Paleozoic, Mesozoic and Cenozoic events} 

The summary for the Anthropocene fingerprint above suggests that similarities  
might be found in (geologically) abrupt events with a multi-variate 
signature. In this section we review a partial selection of known events in the 
paleo-record that have some similarities to the hypothesized eventual anthropogenic signature. 
The clearest class of event with such similarities are the hyperthermals, most 
notably the Paleocene-Eocene Thermal Maximum (56 Ma) \citep{McInerneyWing11}, but this also includes smaller 
hyperthermal events, ocean anoxic events in the Cretaceous and Jurassic, and significant
(if less well characterized) events of the Paleozoic. We don't consider of events (such as the K-T 
extinction event, or the Eocene-Oligocene boundary) where there are very clear and distinct causes 
(asteroid impact combined with massive volcanism\citep{Vellekoop.et.al14}, and the onset of Antarctic glaciation\citep{Zachos.et.al01} (likely linked to the opening of Drake Passage\citep{Cristini.et.al12}), respectively). 
There may be more such events in the record but that are not included here simply because they may not 
have been examined in detail, particularly in the pre-Cenozoic.

\subsection{The Paleocene-Eocene Thermal Maximum (PETM)}

The existence of an abrupt spike in carbon and oxygen isotopes near the Paleocene/Eocene 
transition (56 Ma) was first noted by \citet{KennettStott91} and demonstrated to be global 
by \citet{Koch.et.al92}. Since then, more detailed and high resolution analyses on land and 
in the ocean have revealed a fascinating sequence of events lasting 100--200 kyr and 
involving a rapid input (in perhaps less than 5 kyr \citep{KirtlandTurner.et.al17}) of exogenous carbon into the system 
\citep[see review by][]{McInerneyWing11}, possibly related to the intrusion of the North American Igneous Province into organic sediments\citep{Storey.et.al07}. 
Temperatures rose 5--7\deg C (derived from multiple proxies \citep{TripatiElderfield04}), and there was a negative spike in carbon isotopes 
($>$3\permil), and decreased ocean carbonate preservation in the upper ocean. There was an 
increase in kaolinite (clay) in many sediments \citep{Schmitz.et.al01}, indicating greater 
erosion, though evidence for a global increase is mixed. During the PETM 30--50\% of benthic 
foraminiferal taxa became extinct, and it marked the time of an important mammalian 
\citep{Aubry.et.al98} and lizard \citep{Smith09} expansion across North America. 
Additionally, many metal abundances (including V, Zn, Mo, Cr) spiked during the event 
\citep{Soliman.et.al11}.  

\subsection{Eocene events}

In the 6 million years following the PETM, there are a number of smaller, though qualitatively similar, 
hyperthermal events seen in the record \citep{Slotnick.et.al12}. Notably, the Eocene Thermal 
Maximum 2 event (ETM-2), and at least four other peaks are 
characterized by significant negative carbon isotope excursions, warming and relatively high 
sedimentation rates driven by increases in terrigenous input \citep{DOnofrio.et.al16}. 
Arctic conditions during ETM-2 show evidence of warming, lower salinity, and greater anoxia 
\citep{Sluijs.et.al09}. Collectively these events have been denoted Eocene Layers of 
Mysterious Origin (ELMOs)\footnote{While it is tempting to read something into the 
nomenclature of these events, it should be remembered that most things that happened 50 
million years ago will forever remain somewhat mysterious.}.

Around 40 Ma, another abrupt warming event occurs (the Mid-Eocene Climate Optimum (MECO)), again with an accompanying carbon isotope anomaly \citep{BoscoloGalazzo.et.al14}.

\subsection{Cretaceous and Jurassic Ocean Anoxic Events}

First identified by \citet{SchlangerJenkyns76}, ocean anoxic events (OAEs), identified by periods 
of greatly increased organic carbon deposition and laminated black shale deposits, are times 
when significant portions of the ocean (either regionally or globally) became depleted in 
dissolved oxygen, greatly reducing aerobic bacterial activity. There is partial (though not 
ubiquitous) evidence during the larger OAEs for euxinia (when the ocean 
column becomes filled with hydrogen sulfide (H$_2$S)) \citep{MeyerKump08}.

There were three major OAEs in the Cretaceous, the Weissert event (132 Ma)
\citep{Erba.et.al04}, OAE-1a around 120 Ma lasting about 1 Myr and another OAE-2 around 
93 Ma lasting around 0.8 Myr \citep{Kerr98, Li.et.al08,Malinverno.et.al10,Li.et.al17}. At least four other minor episodes 
of organic black shale production are noted in the Cretaceous (the Faraoni event, OAE-1b, 1d 
and OAE-3) but seem to be restricted to the proto-Atlantic region \citep{Takashimi.et.al06, 
Jenkyns10}. At least one similar event occurred in the Jurassic (183 Ma) \citep{Pearce.et.al08}.

The sequence of events during these events have two distinct fingerprints possibly 
associated with the two differing theoretical mechanisms for the events. For example, during 
OAE-1b, there is evidence of strong stratification and a stagnant deep ocean, while for OAE-2, 
the evidence suggests an decrease in stratification, increased upper ocean productivity and an 
expansion of the oxygen minimum zones \citep{Takashimi.et.al06}.

At the onset of the events (fig. 1c), there is often a significant negative excursion in 
$\delta^{13}$C (as in the PETM), followed by a positive recovery during the events themselves as the 
burial of (light) organic carbon increased and compensated for the initial release 
\citep{Jenkyns10,Naafs.et.al16,Mutterlose.et.al14,Kuhnt.et.al11}. Causes have been linked to the 
crustal formation/tectonic activity and enhanced CO$_2$ (or possibly CH$_4$) release, causing global 
warmth \citep{Jenkyns10}. Increased seawater values of $^{87}$Sr/$^{86}$Sr and $^{187}$Os/$^{188}$Os 
suggest increased runoff, greater nutrient supply and consequently higher upper ocean productivity 
\citep{JonesJenkyns01}. Possible hiatuses in some OAE 1a sections are suggestive of an upper ocean 
dissolution event \citep{Bottini.et.al15}.

Other important shifts in geochemical tracers during the OAEs include much lower nitrogen isotope 
ratios ($\delta^{15}$N), increases in metal concentrations (including As, Bi, Cd, Co, Cr, Ni, V) 
\citep{Jenkyns10}. Positive shifts in sulfur isotopes are seen in most OAEs, with a curious 
exception in OAE-1a where the shift is negative \citep{Turchyn.et.al09}. 

\subsection{Early Mesozoic and Late Paleozoic events}

Starting from the Devonian period, there have been several major abrupt events registered in 
terrestrial sections. The sequences of changes and the comprehensiveness of geochemical analyses 
are less well known than for later events, partly due to the lack of existing ocean sediment, but 
these have been identified in multiple locations and are presumed to be global in extent. 

The Late Devonian extinction around 380--360 Ma, was one of the big five mass extinctions. It's 
associated with black shales and ocean anoxia \citep{AlgeoScheckler98}, stretching from the 
Kellwasser events ($\sim$378 Ma) to the Hangenberg event at the Devonian-Carboniferous 
boundary (359 Ma) \citep{Brezinski.et.al09, DeVleeschouwer.et.al13}. 

In the late Carboniferous, around 305 Ma the Pangaean tropical rainforests collapsed 
\citep{Sahney.et.al10}. This was associated with a shift toward drier and cooler climate, and 
possibly a reduction in atmospheric oxygen, leading to extinctions of some mega-fauna. 

Lastly, the end-Permian extinction event (252 Ma) lasted about 60 kyr was 
accompanied by an initial decrease in carbon isotopes (-5--7\permil), significant 
global warming and extensive deforestation and wildfires 
\citep{KrullRetallack00,Shen.et.al11,Burgess.et.al14} associated with widespread ocean anoxia and 
euxinia \citep{WignallTwitchett96}. Pre-event spikes in nickel (Ni) have also been reported 
\citep{Rothman.et.al14}.


\section{Discussion and testable hypotheses}

There are undoubted similarities between previous abrupt events in the geological record 
and the likely Anthropocene signature in the geological record to come. Negative, abrupt $\delta^{13}$C excursions, warmings, and disruptions of the nitrogen cycle are ubiquitous. 
More complex changes in biota, sedimentation and mineralogy are also common. Specifically, 
compared to the hypothesized Anthropocene signature, almost all 
changes found so far for the PETM are of the same sign and comparable magnitude. Some similarities 
would be expected if the main effect during any event was a significant global warming, however 
caused. Furthermore, there is evidence at many of these events that warming was driven by a 
massive input of exogeneous (biogenic) carbon, either as CO$_2$ or CH$_4$. At least since the 
Carboniferous (300--350 Ma), there has been sufficient fossil carbon to fuel an industrial 
civilization comparable to our own and any of these sources could provide the light carbon input. 
However, in many cases this input is contemporaneous to significant episodes of tectonic and/or 
volcanic activity, for instance, the coincidence of crustal formation events with the climate changes 
suggest that the intrusion of basaltic magmas into organic-rich shales and/or petroleum-bearing 
evaporites \citep{Storey.et.al07,Svensen.et.al09, Kravchinsky12} may have released large quantities 
of CO$_2$ or CH$_4$ to the atmosphere. Impacts to warming and/or carbon influx (such as increased 
runoff, erosion etc.) appear to be qualitatively similar whenever in the geological period they 
occur. These changes are thus not sufficient evidence for prior industrial civilizations. 

Current changes appear to be significantly faster than the paleoclimatic events (figure~\ref{stable}), 
but this may be partly due to limitations of chronology in the geological record. Attempts to time the 
length of prior events have used constant sedimentation estimates, or constant-flux markers (e.g. 
$^3$He \citep{McGeeMukhopadhay12}), or orbital chronologies, or supposed annual or seasonal banding 
in the sediment \citep{WrightSchaller13}. The accuracy of these methods suffer when there are large 
changes in sedimentation or hiatuses across these events (which is common), or rely on the imperfect 
identification of regularities with specific astronomical features \citep{PearsonNicholas14, PearsonThomas15}. 
Additionally, bioturbation will often smooth an abrupt event even in a perfectly preserved sedimentary setting.
Thus the ability to detect an event onset of a few centuries (or less) in the record is questionable, 
and so direct isolation of an industrial cause based only on apparent timing is also not conclusive. 

The specific markers of human industrial activity discussed above (plastics, synthetic pollutants, 
increased metal concentrations etc.) are however
a consequence of the specific path human society and technology has taken, and the generality of 
that pathway for other industrial species is totally unknown. Large-scale energy harnessing is 
potentially a more universal indicator, and given the large energy density in carbon-based fossil 
fuel, one might postulate that a light $\delta^{13}$C signal might be a 
common signal. Conceivably, solar, hydro or geothermal energy sources could have been tapped 
preferentially, and that would greatly reduce any geological footprint (as it would ours). 
However any large release of biogenic carbon whether from methane hydrate pools or volcanic intrusions 
into organic rich sediments, will have a similar signal. We therefore have a 
situation where the known unique markers might not be indicative, while the (perhaps) more expected  
markers are not sufficient.   

We are aware that raising the possibility of a prior industrial civilization as a driver 
for events in the geological record might lead to rather unconstrained speculation. One would be 
able to fit any observations to an imagined civilization in ways that would be basically 
unfalsifiable. Thus, care must be taken not to postulate such a cause until actually positive 
evidence is available. The Silurian hypothesis cannot be regarded as likely merely because no 
other valid idea presents itself.

We nonetheless find the above analyses intriguing enough to motivate some additional research.  
Firstly, despite copious existing work on the likely Anthropocene signature, we recommend further 
synthesis and study on the persistence of uniquely industrial byproducts in ocean sediment 
environments. Are there other classes of compounds that will leave unique traces in the sediment 
geochemistry on multi-million year timescales? In particular, will the byproducts of common 
plastics, or organic long-chain synthetics, be detectable? 

Secondly, and this is indeed more speculative, we propose that a deeper exploration of elemental 
and compositional anomalies in extant sediments spanning previous events be performed (although we 
expect that far more information has been obtained about these sections than has been referenced 
here). Oddities in these sections have been looked for previously as potential signals 
of impact events (successfully for the K-T boundary event, not so for any of the events mentioned 
above), ranging from iridium layers, shocked quartz, micro-tectites, magnetites etc. But it may 
be that a new search and new analyses with the Silurian hypothesis in mind might reveal more. 
Anomalous behaviour in the past might be more clearly detectable in proxies normalized by weathering
fluxes or other constant flux proxies in order to highlight times when productivity or metal 
production might have been artificially  enhanced. Thirdly, should any unexplained anomalies be found, 
the question of whether there are candidate species in the fossil record may become more relevant, as 
might questions about their ultimate fate. 

An intriguing hypothesis presents itself should any of the initial releases of light carbon 
described above indeed be related to a prior industrial civilization. As discussed in section 3.3, 
these releases often triggered episodes of ocean anoxia (via increased nutrient supply) causing a 
massive burial of organic matter, which eventually became source strata for further fossil fuels. 
Thus the prior industrial activity would have actually given rise to the potential for future industry 
via their own demise. Large scale anoxia, in effect, might provide a self-limiting but self-
perpetuating feedback of industry on the planet. Alternatively, it may be just be a part of a long 
term episodic natural carbon cycle feedback on tectonically active planets. 

Perhaps unusually, the authors of this paper are not convinced of the correctness of their 
proposed hypothesis. Were it to be true it would have profound implications and not just for 
astrobiology. However most readers do not need to be told that it is always a bad idea to decide 
on the truth or falsity of an idea based on the consequences of it being true. While we strongly doubt 
that any previous industrial civilization existed before our own, asking the question in a formal 
way that articulates explicitly what evidence for such a civilization might look like raises its own 
useful questions related both to astrobiology and to Anthropocene studies. Thus we hope that this 
paper will serve as motivation to improve the constraints on the hypothesis so that in future we may 
be better placed to answer our title question.

\begin{acknowledgements}
No funding has been provided nor sought for this study. We thank Susan Kidwell for being generous with 
her time and helpful discussions, David Naafs and Stuart Robinson for help and pointers to data for 
OAE1a and Chris Reinhard for his thoughtful review. The GISTEMP data in Fig.~\ref{stable}a was from 
\url{https://data.giss.nasa.gov/gistemp} (accessed Jul 15 2017). 
\end{acknowledgements}

\bibliographystyle{apj}

\end{document}